# THE SHOCK-DRIVING CAPABILITY OF A CME INFERRED FROM MULTIWAVELENGTH OBSERVATIONS

**Fithanegest Kassa Dagnew[1,2,3], Nat Gopalswamy[2], Solomon Belay Tessema[1], Ange Cynthia Umuhire[4], Seiji Yashiro[2,3], Pertti Mäkelä[2,3], and Hong Xie[2,3]**

[1]Ethiopian Space Science and Technology Institute (ESSTI), Entoto Observatory and Research Center (EORC), Addis Ababa, Ethiopia

[2] NASA Goddard Space Flight Center, Greenbelt, MD, USA

[3]The Catholic University of America, Washington DC, USA

[4]University of Rwanda, College of Science and Technology

e-mail: fithanegest.k.dagnew@nasa.gov

## ABSTRACT

The radial speed of a coronal mass ejection (CME) determines the shock-driving capability of a CME as indicated by the presence of a type II radio burst. Here we report on the April 18, 2014 CME that was associated with a type II radio burst in the metric and interplanetary domains. We used the radio-burst data provided by the San Vito Solar Observatory of the Radio Solar Telescope Network and data from the Wind spacecraft. The CME is a full halo in the field of view of the coronagraphs on board the Solar and Heliospheric Observatory (SOHO). The CME was also observed by the coronagraphs on board the Solar Terrestrial Relations Observatory (STEREO). We computed the CME shock and flux rope speeds based on the multi-view observations by the different coronagraphs and by EUV instruments. We determined the shock speed from metric and interplanetary radio observations and found them to be consistent with white-light observations, provided the metric type II burst and its continuation into the decameter-hectometric domain are produced at the shock flanks, where the speed is still high enough to accelerate electrons that produce the type II bursts. Interestingly, there was an interplanetary type II burst segment consistent with an origin at the shock nose suggesting that the curved shock was crossing plasma levels separated by a few solar radii. We conclude that the CME speed is high enough to produce the interplanetary Type II burst and a solar energetic particle (SEP) event. However, the speed is not high enough to produce a ground level enhancement (GLE) event, which requires the shock to form at a height of ~1.5 Rs.

Key words: CME, shock, type II radio burst, SEP event, GLE event.

## 1. Introduction

Type II radio bursts are the earliest indicators of CME-driven shocks (e.g., Gopalswamy 2011). They appear as slow-drifting features at the fundamental and second harmonic of plasma frequency in the radio dynamic spectra. Wild et al. (1963) proposed that the energetic particles might be accelerated at MHD shock waves. Kahler et al. (1978) suggested such shocks are driven by CMEs. Gopalswamy et al. (2008 a, b) suggested the importance of a strong shock in producing a solar energetic particle (SEP) events. Type II solar radio bursts are now considered as the key indicators of SEP events (Gopalswamy et al., 2018). From statistical studies, Gopalswamy (2003) has shown that the majority of large SEP events are associated with type II bursts. Major eruptions with strong X-ray flares and a good magnetic connection to Earth generate large SEP events in the vicinity of Earth (Gopalswamy et al., 2014a). One of the difficulties in understanding the shock-driving capability of CMEs is the complex nature of the type II emission in the interplanetary medium because type II emission can originate from anywhere on the shock surface. Shocks are normally stronger at the noses and weaker on the flanks. CME speeds at the noses are faster than those at the flanks. However, the magnetosonic speeds that determine the shock strength can be substantially different in the nose and flank regions of the shock because the nose is generally at a larger heliocentric distance than the flanks.

In this paper, we exploit the availability of multi-wavelength, multiview observations of the April 18, 2014 CME event that exhibited all the energetic phenomena to obtain a consistent picture of the shock evolution at the Sun and in the near-Sun interplanetary medium. In particular, the interplanetary type II burst is very complex, although the metric type II burst is very simple fundamental-harmonic pair.

## 2. Observations

### 2.1 White-light and EUV observations

The 2014 April 18 CME is associated with a M7.3 flare from NOAA active region (AR) 2036 located at S20W34. The flare begins at 12:31 UT, peaks at 13:03 UT, and ends at 13:20 UT. (see Fig. 1). The CME is fully observed by the two telescopes C2 and C3 of Large Angle and Spectrometric Coronagraph (LASCO, Brueckner et al. 1995) on board the Solar and Heliospheric Observatory (SOHO). The CME is listed in the SOHO/LASCO CME catalog with a first appearance time of 13:25:51 UT in the LASCO/C2 field of view (FOV) and an average speed of ~1203 km/s (Gopalswamy et al. 2009a). The CME is heading in the southwest direction along position angle (PA) of $238^0$, but in the next frame (13:36 UT) it becomes a full halo. The main body of the CME (flux rope) is still in the southwest direction, but the shock extension makes it a full halo. The CME was also observed by the Sun-Earth Connection Coronal and Heliospheric Investigation (SECCHI, Howard et al. 2008): inner coronagraph (COR1), outer coronagraph (COR2), and the Extreme Ultra Violet Imager (EUVI) on board the Solar Terrestrial Relations Observatory (STEREO) spacecraft. The solar source of the CME is also identified in the EUV images obtained by the Atmospheric Imaging Assembly (AIA, Lemen et al. 2012) on board the Solar Dynamics Observatory (SDO) satellite. Figure 1 shows the solar source of the CME as observed by SDO/AIA at 94 Å and the associated GOES soft X-ray flare. STEREO Ahead (STA) and STEREO Behind (STB) were located at W156 and E165, respectively at the time of the eruption (https://stereo-ssc.nascom.nasa.gov/where.shtml). In STA view, the eruption location is E122, which is $32^0$ behind the east limb. Therefore, the CME speed and width can be measured with minimal projection effects. On the other hand, the source is at W199 in STB view, which is directly on the backside. Accordingly, the CME is observed as a backside halo.

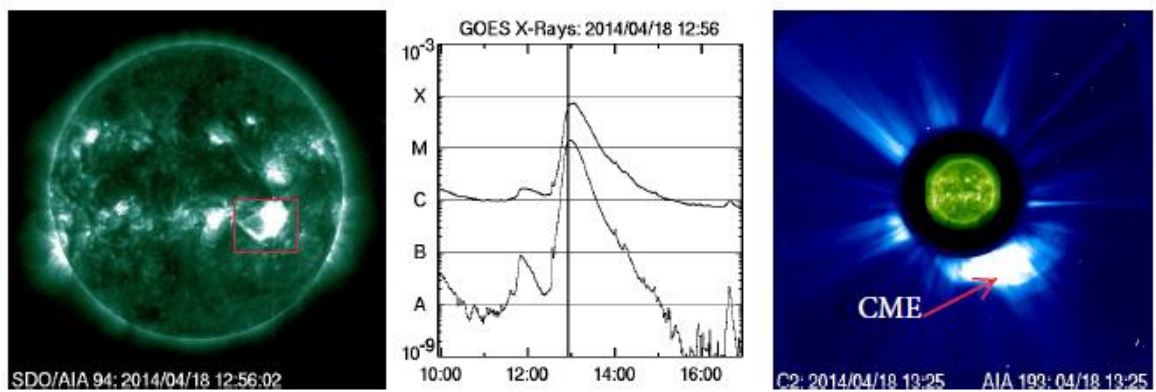

Figure 1: (left) The solar source of the 2014 April 18 eruption as observed in the SDO/AIA 94 Å image obtained at 12:56:02 UT. The post eruption arcade (PEA) at S20W34 is shown enclosed

by a red square. (middle) The GOES soft X-ray light curve in the 1-8 Å (upper) and 0.4 to 5 Å channels. The vertical line marks the time of the AIA image. (right) The CME when it first appeared in the LASCO/C2 FOV at 13:25 UT. The LASCO image is superposed with an AIA/SDO 193 Å image, which also shows the PEA as a bright feature as in the 94 Å image.

Figure 2 shows the CME observed in three views: STA/COR2, SOHO/LASCO, and STB/COR2. In all three views, the CME is observed as a halo because of the source location with respect to the observing spacecraft and most likely the anomalous expansion of CMEs in solar cycle 24 caused by the reduced heliospheric pressure (Gopalswamy et al. 2014b). The CME is roughly directed toward LASCO, while directly anti-sunward in STB view. The CME was also back-sided in STA view as noted before. We use the three-view observations to determine the CME kinematics (see below).

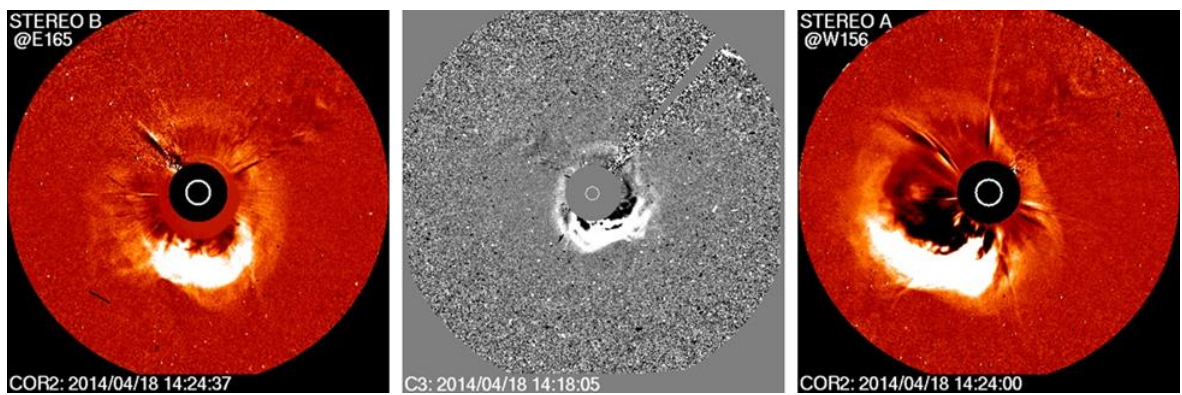

Figure 2: The 2014 April 18 CME as observed by STB/COR2 (left), SOHO/LASCO (middle) and STA/COR2 (right). The locations of STA and STB are given on the plot. The images roughly correspond to the same time. The shock structure can be seen distinctly in STA and STB extending beyond the flux rope (the bright portion of the CME).

**2.2 Radio observations**

The San Vito Solar Observatory of the Radio Solar Telescope Network (RSTN) observed the meter-wave type II radio burst associated with the 2014 April 18 eruption. The radio dynamic spectrum is used to identify the fundamental and harmonic emission of the type II burst. The starting frequency of the fundamental component is 63 MHz with an onset time at 12:55 UT. The burst drifted to the bottom of the dynamic spectrum and can be observed down to ~25 MHz at ~13:03 UT.

The type II burst was also observed by the Radio and Plasma Wave Experiment (WAVES) on board the Wind spacecraft in the interplanetary medium. The type II burst started at 14 MHz and drifted down to 150 KHz (https://cdaw.gsfc.nasa.gov/CME_list/radio/waves_type2.html). The frequency extent suggests that the underlying shock is very strong because it has emission components from metric to kilometric wavelengths (Gopalswamy et al. 2005). In the WAVES dynamic spectrum, the type II burst is a lot more complex with three different segments marked B, C, D (see Fig. 3). Of these, B is a continuation of the metric type II burst into the decameter-hectometric (DH) domain. C starts at the same time as B, but at much lower frequencies. The segment D continues to much lower frequencies, which will be further investigated and reported elsewhere. Here we focus on the segments A, B, and C to understand the relation between the CME and the shock represented by the type II episodes.

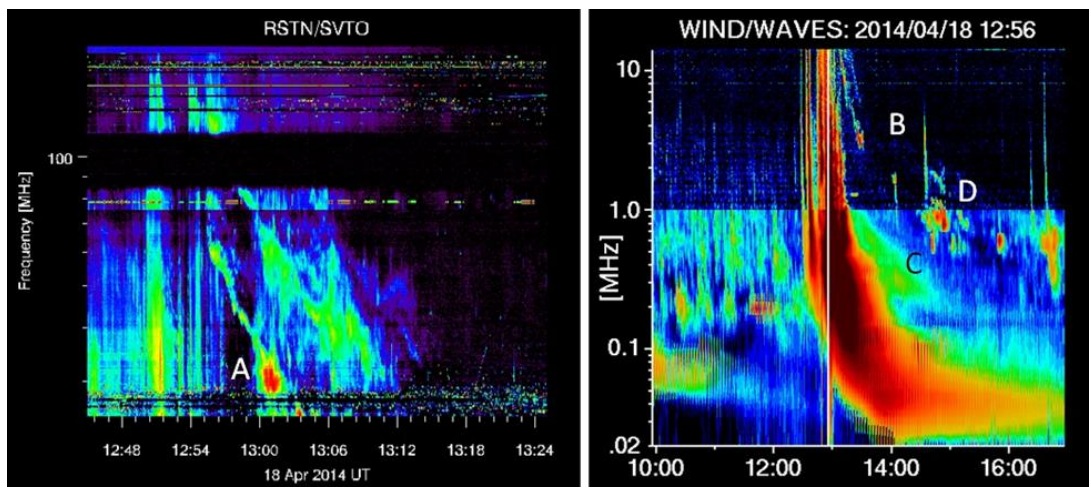

Figure 3: The dynamic spectrum from RSTN (left) and from Wind/WAVES (right). The metric component is marked A in the left figure. The metric type II continues as B in the DH domain lasting until about 3 MHz. A broadband component C appears at the same time as B at frequencies below 1 MHz. In this study, we consider type II emission features at frequencies down to ~3 MHz. Finally, the component D represents the long duration component, which extends to ~150 kHz.

**2.3 SEP observations**

The 2014 April 18 eruption is associated with a large SEP event (>10 MeV proton intensity exceeded 10 pfu; 1 pfu = 1 p cm$^{-2}$ sr$^{-1}$ s$^{-1}$). Figure 4 shows that the proton intensity is significant in the >50 and >100 MeV integral channels also. The SEP event starts in the >10 MeV channel

5 | P a g e

at ~13:40 UT, almost an hour after the onset of the eruption. The >10 MeV proton intensity reaches the 10 pfu level at ~15:00 UT and attains a peak value of ~50 pfu by the end of the day. The delay is due to the tens of minutes taken by the protons to reach the observing satellite and ~10 minutes to accelerate the particles before they are released (Gopalswamy et al. 2012). The eruption is well connected to the GOES satellite, so we observe a sharp increase to the maximum intensity level. Higher energies peak earlier indicating velocity dispersion. The identification of the associated CME and flare is rather unambiguous in this event.

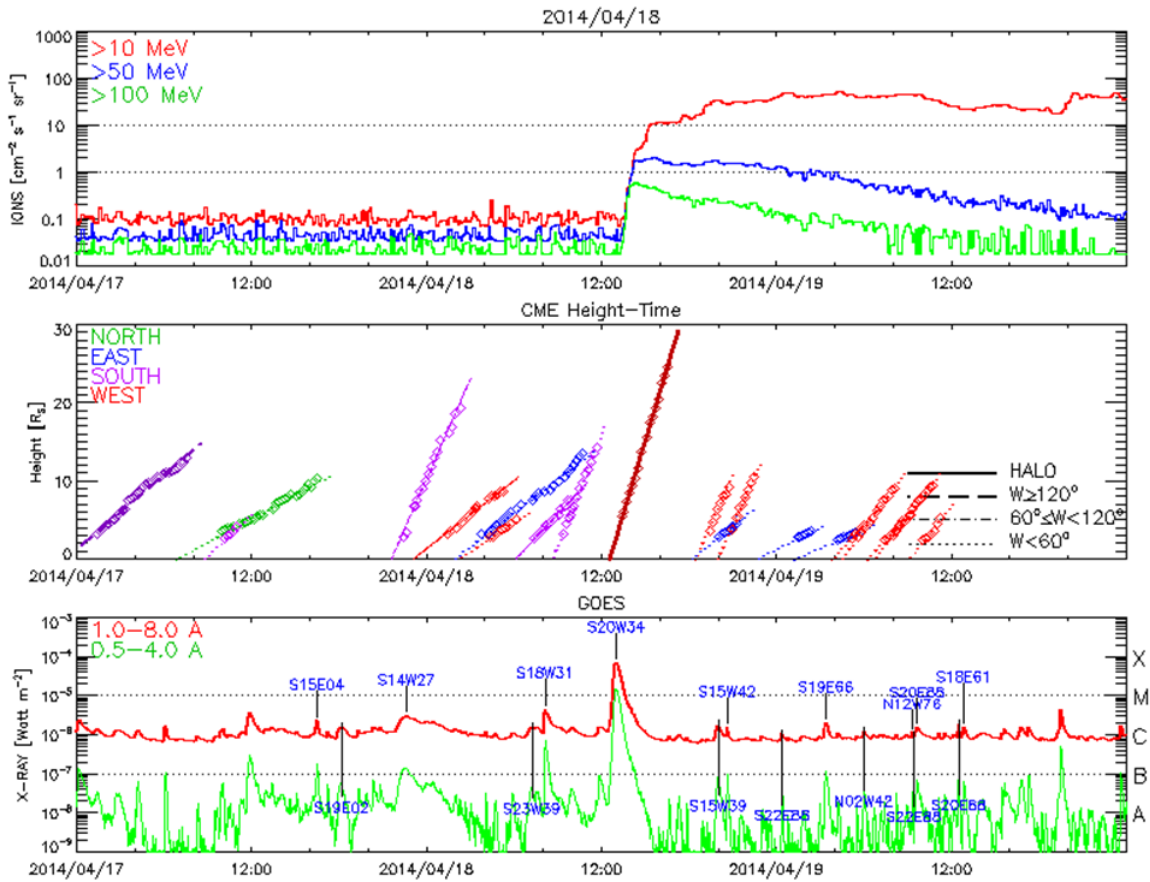

Figure 4: The large SEP event associated with the 2014 April 18 eruption. (top) Proton intensity at >10 MeV, >50 MeV, and >100 MeV energy channels. (middle) CME height-time plots for various CMEs before and after the CME in question starting after 12:00 UT on 2014 April 18. This is the only halo CME in the interval plotted. (bottom) GOES soft X-ray intensity in the 1-8 Å and 0.5 to 4 Å energy channels as a function of time with locations of flares marked. The CME in question is associated with the M-class flare from S20W34.

## 3. Analysis and Results

The CME is observed in three views, so we apply the graduated cylindrical shell (GCS; Thernisien et al. 2009) plus spheroid shock model (Kwon and Vourlidas 2017; Xie et al., 2017) to determine the CME shock kinematics. The GCS model fit is used to determine the three-dimensional CME shape, orientation, radial distance and speed, which is constrained by CME components such as three-part flux-rope like structure or bright frontal loops in the white light (WL) coronagraph images. The spheroid shock model fit is used to describe the shock front, its radial distance and speed, which is constrained by EUV waves and wave-like disturbances in the EUVI and WL images. The fitted shock height-time results give the time variation of the shock speed and acceleration, which we are interested in. We then compare the fitted shock speed to the shock speed derived from the radio dynamic spectrum for the three type II segments, A, B, and C.

Figure 5 shows the leading-edge height, speed and acceleration of the CME driven shock as a function of time. The height-time history is determined by 3 EUVI data points, 3 COR1 data points and 7 COR2 data points. The images used in the height-time plot were taken nearly simultaneously in STEREO and SOHO. The GCS fitted CME propagation direction in COR1 FOV is S20W35, which is similar to the flare location (S20W34). The height-time plot shows that the CME shock accelerates within the COR1 FOV, reaching a peak acceleration of ~0.9 km s$^{-2}$ at ~13:00 UT. The CME shock reaches its peak speed in the next 5 minutes or so and then reaches a steady speed with a slight increase within the COR2 FOV (9.5 m s$^{-2}$). The steady speed is around 1400 km/s. We have also shown the GOES soft X-ray light curve for reference. The initial acceleration peaks before the GOES flare peak as expected. The leading edge is already beyond 2 Rs when the type II burst started in the metric domain and beyond 3 Rs when the burst appeared in the Wind/WAVES spectral domain.

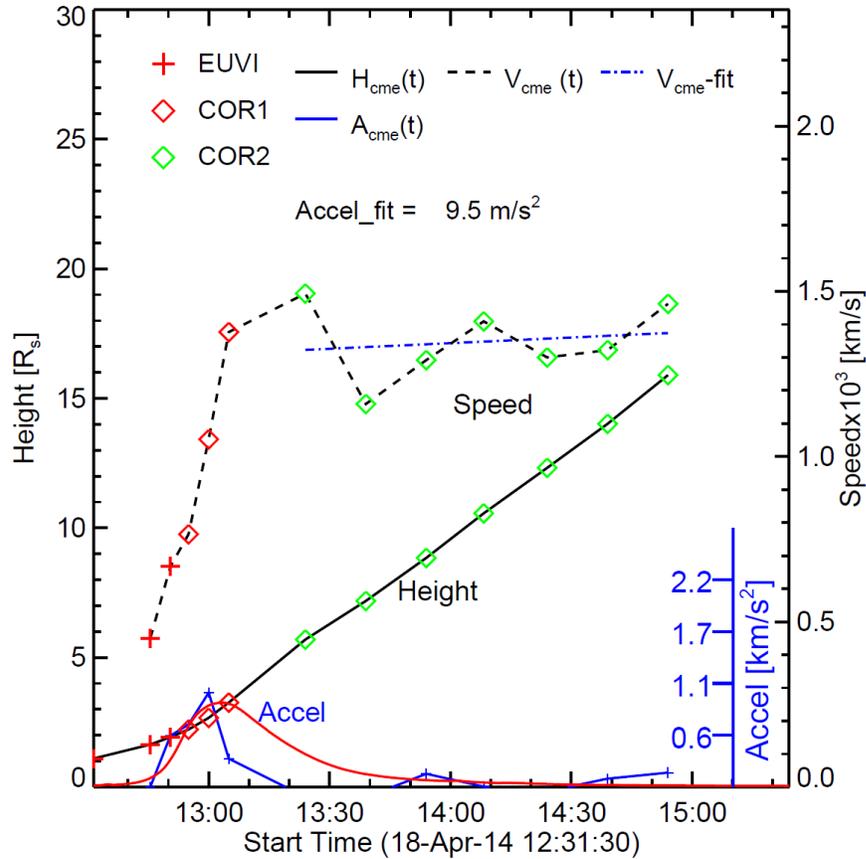

figure 5: Kinematics of the 2014 April 18 CME driven shock based on the cylindrical shell plus shock model fit to data points from STEREO EUVI (red cross), COR1 (red diamond) and COR2 (green diamond) observations: the evolution of CME shock speeds (black dashed line) and the acceleration of shock (blue solid line) derived from the height-time data points, superimposed with the X-ray flare flux (red solid line). The blue dot-dashed line is the linear fit speed of shock for COR2 data points.

### 3.1 The shock speed from metric type II burst

The shock speed is related to the drift rate (df/dt) of the type II burst and the density variation away from the Sun (e.g., Gopalswamy 2011)

$$v_{shock} = (2r/\alpha)\,(1/f_p)\,(df/dt) \quad \ldots\ldots\ldots\ldots\ldots\ldots (1)$$

Here f is the emission frequency, identified with the local plasma frequency ($f_p$); r is the heliocentric distance of the shock at the time of the radio burst (here at the starting frequency);

The drift rate is measured from the dynamic spectrum using the online tool available at the CDAW data center (https://cdaw.gsfc.nasa.gov/) as ~ 0.093 MHz/s; α is the exponent of the density dependence. The electron density variation falls off with the distance (r) from the Sun where the radio emission occurs according to the power law:

$$n(r) \propto r^{-\alpha} \quad \ldots\ldots\ldots\ldots\ldots\ldots\ldots (2)$$

Where α is the slope in the variation of the density over the heliocentric distance. We need to obtain r and α to get $v_{shock}$.

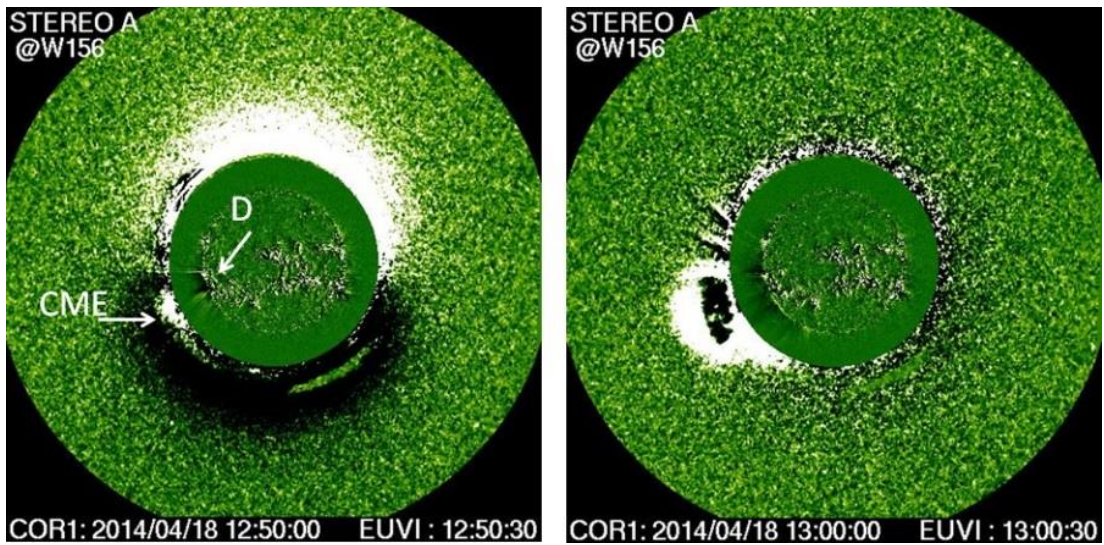

Figure 6. The 2014 April 18 CME before the start of the type II burst at 12:50 UT (left) and just before the end (right) at 13:00 UT. The EUV dimming signature shows that the eruption was behind the east limb STA view. The height of the CME at these times is 1.85 and 2.54 Rs, respectively.

STA/COR1 has the capability of observing CMEs near the onset of metric type II bursts (Gopalswamy et al. 2009b; Gopalswamy et al. 2013a). The CME appeared in the STA/ COR1 FOV at 12:50 UT at r = 1.85 Rs. At this time, the type II burst has not started, indicating that the shock is not formed or it is still too weak to accelerate electrons needed to produce the radio emission. At 13:00 UT, the height is 2.54 Rs suggesting that the local leading-edge speed is ~800 km/s. Taking account of the fact that the eruption is ~$32^0$ behind the limb, the deprojected speed is 944 km/s and the deprojected heights are 2.18 Rs and 3.0 Rs, respectively at 12:50 and 13:00 UT. Actually there is a STA/COR1 image at 12:55 UT, but the leading edge is difficult to determine, so we do not use this image. We extrapolate the CME height obtained at 12:50 UT and 13:00 UT to the time of the type II onset (12:55 UT) as 2.57 Rs. Taking the CME leading edge height as the shock height, we get the shock formation height 2.57 Rs.

Leblanc et al., 1998, Saito (1970), and Newkirk (1967) had worked on density models in the corona and interplanetary (IP) medium. The alpha values vary with the heliocentric distance. The Leblanc, Dulk, and Bougeret (1998) model describes the electron density variation from the low corona to 1AU in an empirical relationship. In the corona from about 1.3 to 3 Rs., coronal observations have shown that the electron density varies proportional to $r^{-6}$. In the interplanetary space, measurements of electron density imply that the radial variation of the electron density is proportional to $r^{-2}$ which is dominant from a few tens of Rs to beyond 1 AU (Leblanc et al., 1998). Hence, in the inner corona where the density falls off sharply, α=6; in the IP medium, α=2; and in the outer corona, α=4.

Although $α = 2$ In the IP medium, the coronal density where the metric type II bursts occur undergoes a sharp decline with distance and the alpha value is taken as $α = 6$ (Gopalswamy et al., 2009 b, Gopalswamy 2011). Using $α = 6$ and r = 2.57 Rs, the shock speed can be computed as 880 km/s, which is consistent with the speed derived from STA/COR1 observations. This speed is also consistent with the three-dimensional speed derived from the GCS fit shown in Fig. 5. However, there is a problem with the height of the 63 MHz plasma level, where the local plasma density is $4.9×10^7$ cm$^{-3}$. For example, statistical results presented by Gopalswamy et al. (2013b) indicate that a starting frequency of 63 MHz corresponds to a heliocentric distance of ~1.5 Rs. This is possible if the radio emission originates from the shock flanks, which are at a lower height compared to the nose. If we consider flanks that make an angle of ~$50^0$ with the nose, the flank height is ~1.65 Rs. Substituting r=1.65 in eq. (1), we get a speed of ~ 565 km/s. If the flank speed decreases from the nose toward the flanks, we can also obtain the flank speed from the nose speed (944 km/s) as 607 km/s, which is not too different from the speed obtained from the type II drift rate. This speed is reasonable because the magnetosonic speed around 1.65 Rs is close to ~400 km/s (Gopalswamy et al. 2001), so the CME is capable of driving a shock required for the type II burst. Gopalswamy (2006) has shown that the typical CME speed at the time of metric type II bursts is ~600 km/s. Therefore, we conclude that the metric type II burst originates from the flanks of the CME. Cho et al. (2007) reported that the metric type II burst arises at the flanks of the CME interacting with the adjacent streamer. As shown in the right panel of figure 6, the metric type II burst originates at the place where the CME interacts with the streamer at the northern flank. Figure 6 shows that the CME is rather small in size at the time of the metric type II burst.

### 3.2 Shock speed in the outer corona and interplanetary medium

From the Wind/WAVES dynamic spectrum, we obtained the drift rate of the fundamental component between 13:19:26 UT and 13:30:24 UT as 0.004 MHz/s at an average frequency of 4.3 MHz. According to the GCS model shown in Fig.5, the shock height at 13:25 UT is ~5.7 Rs and the speed is ~1322 km/s. The plasma frequency of 4.3 MHz typically occurs at ~4 Rs, compared to 5.7 Rs for the height of the shock nose. This can be reconciled when the emission comes from the shock flanks as does the metric emission. This makes sense because the DH emission is a continuation of the metric type II burst. As before, if we consider the flank that makes an angle of ~$50^0$, the flank speed can be obtained from the nose speed (1322 km/s) as ~850 km/s. To get this speed from equation (1), we substitute r = $5.7\cos50^0$ = 3.7 Rs, fp = 4.3 MHz, and df/dt = 0.004 MHz/s, so that we require α = 5.6. This suggests that the shock flank emitting the m-DH type II burst is passing through a structure in which the density is still rapidly declining. If one uses α=2 appropriate to the interplanetary medium, we get a speed of ~2380 km/s, which is about two times the nose speed and hence not plausible. We therefore conclude that the type II emission above 3 MHz is predominantly from the shock flanks.

### 3.3 Type II burst from the shock nose

In addition to the flank emission discussed above, there is a broadband emission starting at ~1 MHz at the same time as the 14 MHz type II burst discussed above. This emission is consistent with the nose emission because frequencies below 1 MHz correspond to larger distances from the Sun. The low frequency component drifts from ~1 MHz to 0.2 MHz and lasts until ~15:00 UT. The drift rate of this component is ~$1.4\times10^{-4}$ MHz/s. A reference frequency of ~0.56 MHz at 13:41 UT corresponds to a CME leading edge height of 7.64 Rs. This distance is far enough into the interplanetary medium, so we can use α=2, so we can get the shock speed from equation (1) as 1329 km/s, which is in close agreement with the CME leading edge speed (1332 km/s) obtained from the GCS model.

### 3.4 The SEP event

The CME speed in this event is close to the average speed (~1500 km/s) of CMEs producing large SEP events and consistent with the fact that SEP events are associated with type II bursts in the DH domain (Gopalswamy 2006). The SEP event in question has already been reported as a soft-spectrum event in Gopalswamy et al. (2016). The fluence spectrum of the SEP event has a power-law spectral index of 4.15±0.05 in the 10-100 MeV energy range. Gopalswamy et al. (2016) showed that the initial acceleration (or speed) is closely related to the spectral hardness of the SEP event. For CMEs with low initial acceleration, the average spectral index is ~4.89 (typical of CMEs associated with filament eruptions outside active regions). Even though the 2014 April 18 CME originated from an active region, the acceleration profiles in Fig. 5 show that the acceleration is not very impulsive and reached a peak value of ~0.9 km s$^{-2}$. Therefore, the spectral index of 4.15 is consistent with the low initial acceleration. Similarly, the average speed of ~1400 km/s is similar to the average speed of CMEs resulting in soft-spectrum SEP events (~1200 km/s), but smaller than the speed of regular SEP events (~1800 km/s) and ground level enhancement events (~2300 km/s).

### 4. Discussion and Conclusion

We analyzed the shock-driving capability of the 2014 April 18 CME from the point of view of the type II burst and SEP association. While three-dimensional speed derived from multi-view observations (~1400 km/s) is high enough to drive a shock, the interplanetary type II burst is highly complex with multiple segments not harmonically related. By a detailed comparison between CME/shock kinematics and the drift rates of various type II segments, we were able to develop a consistent picture. The main finding is that the CME-driven shock produced type II bursts simultaneously from the flanks and nose of the shock.

Another key finding is that the SEP event has a soft spectrum, which is typical of CMEs originating in filament regions outside active regions. Even though the CME in question originated from an active region, the acceleration profile is similar to that of filament eruption events and hence resulted in a soft spectrum SEP event. Physically speaking, the soft spectrum results from the fact that the shock formation height is rather large (~2.57 Rs). This is much larger than the shock formation height of ground level enhancement (GLE) events, which is ~1.5

Rs (Reames 2009; Gopalswamy et al. 2012). At larger distances from the Sun, the acceleration efficiency of the shock declines, so the energy of the accelerated particles cannot attain high values. In this sense, the 2014 April 18 event is consistent with the hierarchical relation between CME kinematics and the spectral hardness of the SEP event reported in Gopalswamy et al. (2016). The specific conclusions of this paper can be summarized as follows:

(1) The metric type II burst has a simple fundamental-harmonic structure that can be explained by emission from the shock flank located at a heliocentric distance of ≤1.65 Rs.

(2) The flank emission continues into the DH domain and lasts until the CME reaches several solar radii.

(3) Around the time when the flank emission reaches the DH domain, another type II segment starts at frequencies below ~1 MHz. This segment is consistent with the emission originating from the shock nose.

(4) The density variation with height derived from the flank emission suggests that the flank passes through a medium in which the density falls of rapidly in the outer corona.

(5) Since SEP protons are predominantly accelerated from the shock nose, the SEP event onset is delayed to the time of the low-frequency type II component from the nose.

(6) The soft spectrum of the SEP event is consistent with the particle acceleration starting at several solar radii from the Sun.

(7) The derived shock speed of ~1400 km/s is sufficient to produce interplanetary type II burst and a large SEP event. However, the speed is not sufficiently high to accelerate particles to very high energies.

## Acknowledgments

We acknowledge NASA's open data policy in using SDO, SOHO, STEREO, and Wind data. STEREO is a mission in NASA's Solar Terrestrial Probes program. SOHO is a project of international collaboration between ESA and NASA. We thank NOAA/NGDC for making the GOES soft X-ray proton data available to be included in the SOHO/LASCO CME catalog available at NASA's CDAW Data Center. This work is supported by NASA's Living With a Star program. FD thanks the Ethiopian Space Science and Technology Institute and the Mekelle University for partial financial support. HX was partially supported by NASA HGI grant NNX17AC47G.**References**

Brueckner, G. E., Howard, R. A., Koomen, M. J., et al.: 1995, Sol Phys, 162, 357.

Cho, K.S., Lee, J., Moon, Y.J., Dryer, M., et al.: 2007, Astron. Astrophys. 461, 1121 – 1125.

Gopalswamy, N., Lara, A., Kaiser, M.L., Bougeret, J.-L.: 2001, J. Geophys. Res. 106, 25261.

Gopalswamy, N., Aguilar–Rodriguez, E., Yashiro, S., Nunes, S., Kaiser, M. L., and Howard, R. A.: 2005, J. Geophys. Res. 110, A12S07.

Gopalswamy, N.: 2006, Geophys. Monogr. Ser. 165, p. 207.

Gopalswamy, N., Yashiro, S., Xie, H., Akiyama, S., Aguilar–Rodriguez, E., Kaiser, M. L., Howard, R.A., Bougeret, J.L.: 2008a, ApJ, 674, 560-569.

Gopalswamy, N., Yashiro, S., Akiyama, S., Mäkelä, P., Xie, H., Kaiser M.L., Howard, R. A., Bougeret, J.L.: 2008b, Ann Geo., 26, 3033-3047.

Gopalswamy, N., Yashiro, S., Michalek, G., Stenborg, G., Vourlidas, A., Freeland, S., Howard, R.: 2009a. Earth, Moon, and Planets, 104, 295.

Gopalswamy, N., Thompson, W. T., Davila, J. M., et al.:2009b, Sol. Phys., 259, 227.

Gopalswamy, N.: 2011, In: Rucker, H.O., Kurth, W.S., Louarn, P., Fischer, G. (eds.), Coronal Mass Ejections and Solar Radio Emissions, Planetary, Solar and Heliospheric Radio Emissions (PRE VII), Austrian Acad. Sci. Press, Vienna, 325.

Gopalswamy, N., Xie, H., Yashiro, S., Akiyama, S., Mäkelä, P., Usoskin, I. G.: 2012, Space Sci. Rev. 171, 23.

Gopalswamy, N., Xie, H., & Akiyama, S. et al. 2013a, ApJL, **765**, L30.
**14 |** P a g e